# Dynamic Solution Probability Acceptance within the Flower Pollination Algorithm for t-way Test Suite Generation

Abdullah B. Nasser[1], Kamal Z. Zamli[1] and Bestoun S.Ahmed[2]

[1]Faculty of Computer Systems and Software Engineering,
Universiti Malaysia Pahang, 26300 Kuantan, Pahang, Malaysia

[2]Faculty of Electrical Engineering, Department of Computer Science, Czech Technical University, Czech Republic

**Abstract.** Flower Pollination Algorithm (FPA) is the new breed of meta-heuristic for general optimization problem. In this paper, an improved algorithm based on Flower Pollination Algorithm (FPA), called imFPA, has been proposed. In imFPA, the static selection probability is replaced by the dynamic solution selection probability in order to enhance the diversification and intensification of the overall search process. Experimental adoptions on combinatorial t-way test suite generation problem (where t indicates the interaction strength) show that imFPA produces very competitive results as compared to existing strategies.

## 1      Introduction

Meta-heuristic is higher level of stochastic methods that attempt to escape from local optimum, by applying intelligent concepts of exploring and exploiting search space. Meta-heuristic algorithms have successfully being used for solving combinatorial problems which cannot be solved in one step, which arise in many areas of computer science and software engineering, and its applications including software design, project planning and cost estimation, requirement engineering, network packet routing, protein structure prediction, software measurement and software testing [1-6]. As a result, many meta-heuristic algorithms have been proposed in the literature (such as Tabu Search (TS), Simulated Annealing (SA), Differential Evolution (DE), Genetic Algorithm (GA), Anti Colony Algorithm (ACA), Particle Swarm Optimization (PSO), Harmony Search (HS), Sine Cosine Algorithm (SCA), Bees Algorithm (BA), Cuckoo Search (CS), Bat Algorithm (BA), Teaching–learning-based optimization (TLBO) and Firefly Algorithm(FA), to name a few [7]).

Most of above-mentioned algorithms use parameters to guide and control the local and global parts of search process (via intensification and diversification).Here, intensification explores the promising neighboring regions and diversification to ensure that all regions of the search space have been explored. For example, GA uses crossover, mutation, and selection operators, and TLBO uses teacher and the learner's operator. Complementing existing work, our work focuses on enhancing the exploration and exploitation of the Flower Pollination Algorithm (FPA). We argue that existing work on FPA [8] uses a simple probability $p_a$ to control the exploration (i.e. global pollination) and exploitation (i.e. local pollination). As the search space is dynamic (i.e. based on the given configuration), any fixed and preset $p_a$ can be counter – productive as far as exploration and exploitation is concerned. In fact, there should be more exploration (i.e. via global pollination operator) in the early part of the searching process. Then, towards the end, there should be more exploitation (i.e. via local pollination operator). To do so, FPA needs to do away with its $p_a$ (i.e. probability selection of operators) allowing both global and local pollination to run in sequence. Here, the probability is introduced as a dynamic parameter to select the solution itself (rather than the operators). Initially, the probability is high to accept all solution (for diversification). This probability will decrease with time and only allows good solution to be accepted thereafter. In this paper, we introduce a new variant of FPA called imFPA. As a case study, we adopt the problem from Search based Software Engineering domain involving the t-way combinatorial test suite generation problem (where t represents the interaction strength). Experimental results show that imFPA produces very competitive results as compared to existing strategies.

## 2  Overview of t-way Testing and Related Work

Within t-way testing (where t indicates the interaction strength), all t-combinations of system's components must be covered at least once. As illustration, Fig. 1 shows the window of power and sleep setting in Windows 10.

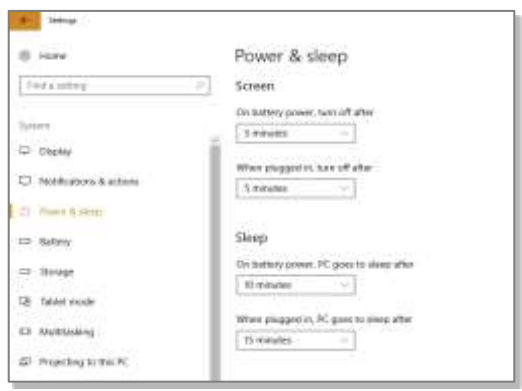

**Fig. 1.** Customize setting for power and sleep in Windows 10

The window consists 4 components (i.e. time to turn off screen on battery power, time to turn off screen when plugged in, time to sleep on battery power and time to sleep when plugged in), each component or input have 16 values (i.e. 1 minuet, 2 minutes, 3 minutes, 5 minutes, 10 minutes, 15 minutes, 20 minutes, 25 minutes, 30 minutes, 42 minutes, 1 hour, 2 hours, 3 hours, 4 hours, 5 hours, and never). Ideally, there is need to test all possible interaction between the inputs' values with 65,536 test cases. However, if only every two pairs (2-way testing) of the components are considered, the final test suite size can be reduced to 310 test cases. Here, the final test suite includes all possible of any two inputs at least once, which can detect %76 of software failure [9], while 3-way testing can reduce the test cases to 5300 test cases..

Recently, many meta-heuristic algorithms have been used for sampling of t-way test suite. In general, meta-heuristic-based strategies start with a random set of solutions, called *population*. The population is subjected to a sequence of search operation in an attempt to improve them. At each iteration, the best candidate solution is selected and added to the final test suite.

GA and ACA [10] are the first population based algorithm has been used for generating t-way testing. GA adopted three operators, selection, crossover and mutation, for diversification and intensification. ACA are probability algorithm controlled by 6 parameters (i.e. pheromone control, pheromone persistence, max stale period, pheromone amount, elite ants, and heuristic controls) apart from population size and number of iteration [10].

PSO have been adopted for generating t-way test suite [11, 12] due to its excellent performance. The search in PSO depends on mathematical formulae, therefore the diversification and intensification is implicitly implemented by updating its particle velocity [13]. Similar to PSO, HS adopts mathematical formulae to move to better solution and uses a probabilistic-gradient to select the current solution. In addition, HS employs transformer and elitism selection for intensification and diversification, respectively [14].

Both CS [15] and FPA[16] use Lévy flight in search process where Lévy flight can be consider as diversification and intensification based on step size of Lévy flight. Addition to Lévy flight, CS uses elitism mechanism to increase the diversity of the population, while FPA uses learning mechanism to intensive the search around each flower (i.e. Local pollination). Bat Algorithm (BA) has been used for generating test suite. The algorithm mimics the hunting behavior of Microbats. The intensification part in is implemented using random walk around the current best, while diversification part have some similarity to PSO [17].

Unlike aforementioned strategies, High Level Hyper-Heuristic (HHH) [18] uses four meta-heuristic algorithms in its implementation include TS (as High level algorithm) and PSO, CS and Global Neighborhood algorithm (as low level algorithms) for generating test suite. Adaptive TLBO (ALTBO) [19] is also used for generating test suite. ATLBO improves the performance of TLBO resulting from a good balance between diversification and intensification through the adoption of fuzzy inference rules. Recently, Q-learning has also been adopted with Sine Cosine Algorithm (QSCA) [20] for t-way test suite generation.

## 3 Proposed Strategy

Standard FPA starts by generating a number of random population, then applies either global pollination or local pollination, controlled by switch probability $p \in [0, 1]$, on the population [21]. . The complete pseudocode of imFPA is presented Fig. 2.

Unlike FPA, imFPA adopt a new mechanism for selecting local and global pollination in order to improve the performance of FPA. In imFPA, maintaining the characteristics of flower pollination, the new solutions generating using global pollination or local pollination will be accepted if it better than current solutions, if not it will be accepted with probability P(f(x), f($x_{new}$), success_rate) or accepted with new. By doing so, the selection mechanism considers the improvement of current solutions and the total success of global and local pollinations.

```
Input:  (P, v, t):   Set of parameters P, its parameter-value v and interaction strength.
Output:    TS final test cases
Generating Interaction tuples
    Let TS be a set of final test cases;
    Generate initial population of Pop_min pollen randomly
    while Interaction tuples are not cover do
        while t < MaxGeneration or stop criterion do
            for ( each x_t^i in the population)
                Global pollination via x_t^{i+1} = x_t^i + L(gbest − x_t^i)
                if (x_t^{i+1} better than x_t^i )
                    Accept x_t^{i+1}
                    Global_success_rate++
                else accept x_t^{i+1} with probability P(f(x), f(x_new), Global_success_rate)
                end if
                Randomly choose j and k among all the solutions
                Do local pollination via x_t^{i+1} = x_t^i + ϱ(x_t^j − x_t^k)
                Global pollination via x_t^{i+1} = x_t^i + L(gbest − x_t^i)
                if (x_t^{i+1} better than x_t^i )
                    accept x_t^{i+1}
                    Local_success_rate++
                else accept with probability P(f(x), f(x_new), Local_success_rate)
                end if
            end for
            Find the current best pollen gbest
        end while
        Add the best pollen gbest into TS.
        Remove covered interactions elements from IE.
    end while
```

**Fig. 2.** imFPA strategy for t-way test cases generation

## 4 Results and Evaluation

In this section, imFPA is compared with existing strategies - GA, SA, PSO, CS, HS and HHH. Different system configurations are used as shown in Table 1, first and second columns, where $x^y$ indicates that the system configuration has y parameters, each parameter has x values, and $t$ is required interaction. Each cell in Table 1 displays the best test case obtained by each strategy, while the dark cell displays the best minimum size obtained among the existing strategies. For experimental setup, the values of population size = 500, and iteration = 500 have been used. Each configuration is run 30 times, and the best result obtained is recorded.

Table 1 depicts that the results of imFPA are very competitive with the existing strategies. imFPA obtains the most minimum results five cases (S1, S3, S8, S11 and S14), while HHH obtains the best six times. In general, we can observe that meta-heuristic based strategies, such as HS, CS, HHH, FPA and imFPA, perform better than other strategies. In order to investigate the behavior of FPA and imFPA, global and local pollination for both FPA and imFPA have been tracked. As Fig.3 shows, the mean percentage of global and local imFPA are changed based on the problem, however, in FPA, it remains the same for all problem instances.

**Table 1.** Comparison of imFPA with existing strategies

| Systems | | | IPOG | ITCH | Jenny | PICT | TVG | GTWay | SA | GA | ACA | PSO | HSS | HHH | CS | FPA | imFPA |
|---|---|---|---|---|---|---|---|---|---|---|---|---|---|---|---|---|---|
| No | $x^y$ | $t$ | | | | | | | | | | | | | | | |
| S1 | $3^4$ | 2 | 12 | NA | 10 | 13 | 12 | 10 | 9 | 9 | 9 | 9 | 9 | 9 | 9 | 9 | 9 |
| S2 | $3^{13}$ | 2 | 20 | NA | 22 | 20 | 20 | 19 | 16 | 17 | 17 | 17 | 18 | NA | NA | 18 | 18 |
| S3 | $10^{10}$ | 2 | 176 | NA | 177 | 170 | 189 | 160 | NA | 157 | 159 | 170 | 155 | NA | NA | 153 | 151 |
| S4 | $5^5$ | 4 | 908 | 837 | 810 | 773 | 731 | 625 | NA | NA | NA | 779 | 751 | 746 | 776 | 784 | 790 |
| S5 | $5^6$ | 4 | 1239 | 1074 | 1072 | 1092 | 1027 | 625 | NA | NA | NA | 1001 | 990 | 967 | 991 | 988 | 988 |
| S6 | $5^7$ | 4 | 1349 | 1248 | 1279 | 1320 | 1216 | 1125 | NA | NA | NA | 1209 | 1186 | 1151 | 1200 | 1164 | 1165 |
| S7 | $2^{10}$ | 4 | 49 | 58 | 39 | 43 | 40 | 46 | NA | NA | NA | 34 | 37 | 36 | 28 | 36 | 36 |
| S8 | $3^{10}$ | 4 | 241 | 336 | 221 | 231 | 228 | 224 | NA | NA | NA | 213 | 211 | 207 | 211 | 211 | 205 |
| S9 | $4^{10}$ | 4 | 707 | 704 | 703 | 742 | 782 | 621 | NA | NA | NA | 685 | 691 | 668 | 698 | 661 | 657 |
| S10 | $2^{10}$ | 2 | 10 | 6 | 10 | NA | 10 | NA | NA | NA | NA | 8 | 7 | 8 | 8 | 8 | 8 |
| S11 | $2^{10}$ | 3 | 19 | 18 | 18 | NA | 17 | NA | NA | NA | NA | 17 | 16 | 16 | 16 | 16 | 16 |
| S12 | $2^{10}$ | 4 | 49 | 58 | 39 | NA | 41 | NA | NA | NA | NA | 37 | 37 | 36 | 36 | 35 | 37 |
| S13 | $2^{10}$ | 5 | 128 | NA | 87 | NA | 84 | NA | NA | NA | NA | 82 | 81 | 79 | 79 | 81 | 82 |
| S14 | $2^{10}$ | 6 | 352 | NA | 169 | NA | 168 | NA | NA | NA | NA | 158 | 158 | 153 | 157 | 158 | 153 |

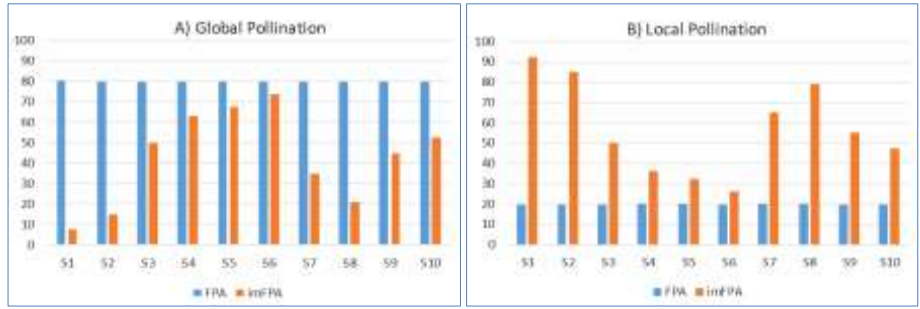

**Fig. 3.** Mean of global and local search percentage of FPA and imFPA

## 5    Conclusion

In this paper, an improved algorithm based on Flower Pollination Algorithm (FPA), called imFPA, is proposed. imFPA provides a new mechanism for candidate solution's selection. The experimental results show that imFPA produces competitive results comparing with existing strategies. As part of the future work, we are currently investigating imFPA for constraints software product lines test suite generation.

### Acknowledgements

This work is funded by FRGS Grant from the Ministry of Higher Education Malaysia titled: A Reinforcement Learning Sine Cosine based Strategy for Combinatorial Test Suite Generation (grant no: RDU170103).